# Working Paper Number 166

# Local Knowledge and Natural Resource Management in a Peasant Farming Community Facing Rapid Change: A Critical Examination

## Jules R. Siedenburg[1]

*Environmental degradation is a major global problem, and addressing it is a key Millennium Development Goal. Its impacts are not just environmental (e.g., species loss), but also economic (e.g., reduced agricultural productivity), with degradation increasingly cited as a key cause of rural poverty in the developing world.*

*The degradation literature typically emphasises common property or 'open access' natural resources, and how perverse incentives or missing institutions lead optimising private actors to degrade them. By contrast, the present paper considers degradation occurring on private farms in peasant communities. This is a critical yet delicate issue, given the poverty of such areas and questions about the role of farmers in either degrading or regenerating rural lands*

*The paper examines natural resource management by peasant farmers in rural Tanzania. Its key concern is how the local knowledge informing farmers' management decisions adapts to challenges associated with environmental degradation and market liberalisation. Given their poverty, this question could have direct implications for the capacity of households to successfully meet their livelihood needs.*

*Based on fresh empirical data, the paper finds that differential farmer knowledge helps explain the large differences in how households and communities respond to the degradation challenge. The implication is that some farmers adapt more effectively to emerging challenges than others, despite all being rational, optimising agents who follow the management strategies they deem best. The paper thus provides a critique of local knowledge, implying that some farmers experience adaptation slippages while others race ahead with effective adaptations.*

*The paper speaks to the chronic poverty that plagues many rural communities in the developing world. Specifically, it helps explain the failure of proven 'sustainable agriculture' technologies to disseminate readily beyond an initial group of early innovators, and suggests a means to help 'scale up' local successes. Its key policy implication is to inform improved capacity building for peasant communities.*

**April 2008**

[1] Jules Siedenburg is a freelance consultant working in the area of climate change – development linkages in the agriculture, forestry and bioenergy sectors, notably for Winrock International and Oxfam America. This paper draws upon a study conducted for the author's doctoral degree at QEH in collaboration with ICRAF's Tanzania country office. The author is grateful to ICRAF, his field staff and participating farm households and local experts for their generous and thoughtful contributions to this study.

*Introduction & hypothesis*
*i. Anecdotes from the field*

During the two years I spent living in a remote village in rural Senegal, I observed major differences in how local farmers managed natural resources (NRs). Notably, tree management strategies varied sharply among seemingly similar farms, with some households removing all trees, some retaining certain naturally-occurring species, and some actively cultivating selected species in particular configurations.

When during visits to farms I asked why a given neighbour followed a different strategy, a common response was for the farmer to lean forward and whisper conspiratorially, "he doesn't know what he's doing!" When I later spoke with said neighbour, he often made the same assertion in reverse. Had the views expressed merely concerned personal preferences or opinions, this would have been amusing but inconsequential commentary on human nature. Yet a second anecdote shows the major welfare implications of these different strategies.

One year the rainy season proved erratic. The rains began early with a powerful storm that caused the pre-sown millet crop to sprout, after which it did not rain again for five weeks. When the rains finally resumed, a large majority of the millet fields had to be replanted, the original crop having died. Such fields ultimately produced little, given the brevity of their growing season, imposing hardship on many households. A minority of fields survived, however, and their crops grew large and produced abundantly. Based on the author's straw poll, these farms were distinguished by management practices such as retaining or cultivating farm trees, manuring, and mulching rather than burning crop residues. That is, they were distinguished by 'regenerative practices' that foster outcomes such as higher levels of soil organic matter and a more clement microclimate.

The implication seemed to be that some ideas about natural resource management (NRM) – and hence some management strategies followed – were 'better' adapted to the contemporary local context than others, despite each farmer doing what he/she thinks best. That is, some farmers may follow strategies that are not fully adapted to contemporary local constraints, needlessly making due with reduced welfare despite being rational, optimising and familiar with the local environment.

This conundrum ultimately led to my doctoral work, which critically examined the 'local knowledge' (LK) informing NRM by peasant farmers in a district of Tanzania experiencing rapid population growth and environmental degradation. LK is the knowledge that informs the day-to-day decisions of peasant farmers, including their NRM decisions. It is derived primarily from farmers' observations and historical experience, but is also influenced by 'outside' information sources such as schooling, extension services and radio.

Looking ahead, the study's findings confirm the preceding thumbnail interpretation of the Senegal anecdote. That is, they suggest that gaps within LK vis-à-vis NRM may critically constrain rural livelihoods under certain circumstances, while addressing these gaps offers a promising policy lever for poverty reduction efforts.

The present paper first makes introductory observations about contemporary peasant agriculture, then presents empirical evidence based on an in-depth household survey conducted in Shinyanga Rural District, Tanzania. The subsequent discussion seeks to understand the phenomenon of observed discrepancies in farmer knowledge vis-à-vis NRM and how they might occur among rational, optimising farmers. The paper concludes by highlighting policy implications, notably for ongoing poverty reduction efforts.

Since it was based on a random sample of a district, the study was most relevant to the 'silent majority' of peasant farmers that do not have regular project or extension contacts but must nonetheless cope with dramatic emerging challenges via their LK, in contrast to the farmers whose learning and adaptive innovations are based on collaborations with outsiders, as documented by the adaptive co-management literature.

It should be also noted that while this analysis focuses on peasant farmers and LK, it could just as easily have explored ways in which rich, industrialised societies may struggle to adapt NRM to growing NR scarcity[2]. Yet NRM by peasant farmers is arguably a more pressing concern, given the poverty of such farmers and their direct reliance on local NRs to meet their basic needs.

*ii. Poverty and the 'peasant farmer'*

Vast rural areas containing hundreds of millions of people remain mired in poverty, notably in South Asia and Sub-Saharan Africa. About 1 billion people still live on less than US$1/day and 2.6 billion on less than US$2/day, while 830 million are hungry. The poor live mainly from farming, notably as 'peasant farmers'[3] engaged in semi-subsistence production (IFAD 2001; HDR 2007; IFPRI 2007). Peasant communities occur primarily in 'low-potential' areas or 'less-favoured' lands[4]. These areas are home to 1.8 billion people worldwide, including a large majority of the rural poor in developing countries (IFPRI 2002). Addressing the challenges facing such communities is thus central to securing poverty reduction, the prime Millennium Development Goal and a key global challenge for the twenty-first century.

Despite diverse efforts over many years to address Tanzania's economic and social problems, poverty in the country remains widespread. Based on the 1992 Household Budget Survey, one-quarter of Tanzanians lived in households unable to meet their nutritional requirements, while about half were unable to meet their combined food and non-food basic requirements. Estimates for the year 2000 suggest that poverty may have increased since then. Contemporary Tanzanian poverty is largely rural, with the poor concentrated in semi-subsistence agriculture on 'low potential' lands (PRSP 2000).

These poverty statistics reflect the serious challenges facing African farmers. Although agriculture is the prime occupation of a large majority of the population in most African countries[5], food production per capita has been falling in recent years (Pinstrup-Anderson et al 1997). In Tanzania, cereal yields per hectare have fallen by 8 percent over the period 1990-2001, while meat production per capita has fallen by 12 percent (World Resources 2002). Despite such trends, agriculture remains the lead sector in most low income countries, partly due to its diverse forward and backward linkages (Pinstrup-Anderson & Pandya-Lorch 2001).

*iii. Natural resource degradation as problem*

Peasant economies face different opportunities and constraints from modern rural economies. Notably, viable NR stocks are critical to rural livelihoods in these areas, which are often 'essentially biomass-based subsistence economies' (Dasgupta 1997). Specifically, "75% of the world's poor are rural poor who depend directly on natural systems for their livelihoods" (World Resources 2005). Farming in these areas depends upon local NRs to provide key inputs, since the rural poor often cannot access 'modern' agricultural inputs such as agrochemicals and hybrid seeds (Upton 1996). Specifically, the local environment provides goods such as fuel, fodder, timber and herbal medicine and services such as soil fertilisation and a stable, clement microclimate, making NRM central to rural livelihoods in these communities. Even where peasant farmers diversify their livelihoods (Bryceson 2002), these alternative occupations (e.g., making charcoal, handicrafts) often involve exploiting NRs, and hence likewise depend on the vitality and ongoing viability of these resources (Belshaw and Livingstone 2002).

Despite the centrality of NRs to rural livelihoods, environmental degradation is widespread in the developing world, including in peasant farming areas. Citing harrowing figures of environmental decline, World Resources 2005 argues that the key MDG of halving poverty by 2015 cannot be met without better environmental stewardship. Environmental degradation is also significant. Notably, it is increasingly cited as a key cause of the poor performance of African agriculture (e.g., World Bank 1996; Lutz 2000; Scherr & Yadav 2001, Koning & Smaling 2005).

---

[2] After all, the broader phenomenon in question – i.e., halting adaptation of knowledge or institutions to emerging NR scarcity, and continued reliance on natural processes to maintain NR stocks or assimilate pollutants – applies to people generally, as exemplified by the US 'Dustbowl' in the 1930s or the growing impacts of contemporary climate change.
[3] Peasant farmers typically rely on household labour to work small farms, producing most of their own food and only sometimes selling to markets. They tend to engage in both crop and livestock production and to rely on locally available NRs to provide key farm inputs and consumables (Ellis 1993).
[4] These are defined as areas that face difficult agro-climatic conditions, such as poor soils, low rainfall, or steep slopes. They may also have limited access to infrastructure such as roads, markets and irrigation works or services such as extension, schools and health centres (IFPRI 2005; Pender et al 2001).
[5] Despite evidence that peasant households are diversifying their livelihoods to minimse risk and increase earnings (Bryceson 2002), most still depend primarily on agriculture for their livelihood (Belshaw & Livingstone 2002).

Clearly, such trends are not universal. Indeed, Vogt et al (2006) note that mechanisms resulting in environmental degradation and enhancement may occur simultaneously, and that farmers may either degrade or regenerate rural lands. These authors call for research to better characterise these dynamics and hence to identify effective policy interventions.

Some authors, however, have argued more broadly against what they see as a degradation bias in the literature. For instance, Leach & Fairhead (1996), Gray (1999), and Tiffen et al (1994) extrapolate from exciting case studies of farmer-led environmental regeneration to question pervading perceptions of widespread environmental degradation in Sub-Saharan Africa, suggesting they reflect a wider 'misreading' of the African landscape. Yet extrapolating from such cases seems ill-advised, given the implication that degradation need not concern policy makers. In the case of Shinyanga Rural District, certainly, degradation remains of great concern, as reflected in both the household survey conducted and the existing literature.

Farm communities that use NRs in ways that allow them to degrade are said to be 'mining' the resource base or depleting NR stocks (e.g., soils, pastures, forests, wildlife, aquifers), which reduces land's productive capacity. Typically, NR 'mining' involves increased use pressures – as reflected in disappearing fallow periods[6] – coupled with failure to actively maintain key resources (e.g., via green manuring or agroforestry practices) or substitute for them (e.g., via chemical fertilisers or kerosene).

Contemporary Shinyanga Rural District seems to fit this pattern. Land is scarce in the district, as reflected in the survey finding of short to non-existent fallow periods. At the same time, conventional soil amendments often fail to maintain fertility. On the one hand, households largely lack access to Green Revolution inputs such as chemical fertilisers, while on the other manure supplies are often insufficient. These factors help explain the current reality of falling yields and scarcity of essential products in the area (HASHI/ICRAF 2000).

Simply put, the peasant production system – whereby livelihoods depend largely on the ongoing vitality of local NRs – is in many cases not working well. In such cases innovations are needed, whether in the form of innovative farming practices or by seeking out an alternative livelihood strategy. Two broad types of agricultural innovation may be distinguished, both involving adoption of innovative practices by farmers. Green Revolution agriculture involves shifting to a reliance on purchased agricultural inputs such as chemical fertilisers and hybrid seeds, i.e., on 'modern' substitutes for traditional natural inputs. Sustainable agriculture, meanwhile, involves actively maintaining stocks of key NRs –e.g., fertile soil, fuel or fodder supplies – which might otherwise become degraded if use pressures are intense.

**Table 1   Peasant farming and competing pathways to agricultural innovation**

*iv. Competing innovations as response:  Green Revolution vs sustainable agriculture*

The Green Revolution began in the 1960s with technological innovations involving hybrid seeds, chemical inputs and irrigation. It has been hugely successful, greatly increasing the production of developing world farms[7], and allaying Malthusian fears about the world's capacity to meet people's food needs by enabling food supply to exceed population growth. Recently others have spoken of a 'doubly green revolution' emphasising genetically modified crops (e.g., Conway & Ruttan 1999).

Despite these successes, use of Green Revolution technologies has been concentrated among better-off farmers in high-potential areas of developing countries (IFPRI 2002; World Bank 1998). Meanwhile, peasant farmers may struggle to afford Green Revolution inputs, particularly since the widespread abolition of input subsidies during the 1980s and 1990s. Even where farmers can access them, Green Revolution technologies may be uneconomical in poor agricultural communities due to factors such as poor infrastructure or low yield response on degraded soils (Hazell & Garrett 1996).

---

[6] Where land is abundant, declining soil fertility can be addressed by leaving fields fallow or moving to a new area (i.e., shifting cultivation). This extensive farming strategy may be optimal where land is ample (Upton 1996). Yet where fallow periods become short due to intensive use pressures on land, fallowing may be inadequate to replenish depleted resources, leading to land degradation.

[7] E.g., in the past forty years, yields per hectare in developing countries have increased 208% for wheat, 109% for rice and 157% for maize (MEA 2005:116).

More generally, purchased inputs may be problematic where farm outputs are consumed directly instead of sold to raise cash income.

By contrast, sustainable agriculture depends upon 'regenerative practices' that are accessible to farmers even in marginal areas. Key subsets of regenerative practices include soil and water conservation[8], agroforestry[9], and integrated pest management[10]. The accessibility of such practices stems from their emphasis on 'natural' inputs, notably on NRs available within rural communities (Altieri et al 2001).

Regenerative practices are labour-intensive, actively maintaining key NRs, as opposed to relying on nature's bounty (as with 'shifting cultivation') or purchases of external inputs (as with Green Revolution agriculture). They facilitate regenerative processes and harness symbioses between farm components such as crops, livestock and trees[11]. Because they involve 'investments' in natural capital, regenerative practices can restore productive potential lost due to environmental degradation, or increase the carrying capacity of land (Pretty et al 1996). As such, they represent a means of meeting growing demands on land due to population pressure (Scherr & Yadav 1996), potentially turning a 'vicious' circle of agricultural stagnation and environmental degradation into a 'virtuous' circle of growth and regeneration (World Bank 2008a). They can also minimise the need for external inputs, thus delivering significant cost savings (Pretty 1995).

It has been suggested that one defining characteristic of sustainable agriculture is that it is information intensive, in contrast to the two 'green revolutions', which are capital intensive (Sillitoe 1998). This follows from its reliance on complex interactions among farm components and active NRM. Knowledge is thus particularly important to farmers who lack access to green revolution technologies (Brokensha et al 1980).

The promise of sustainable agriculture to improve livelihoods in poor rural areas has been amply demonstrated. Based on a multi-country analysis, Pretty et al (1996) report that adoption of regenerative practices such as agroforestry technologies led to large productivity increases in 'low potential', food-deficit areas of developing countries. On average, adopters had doubled their cereal production since adoption in addition to diversifying farm outputs, notably through production of useful tree products. Similarly impressive results have been obtained in Shinyanga District, Tanzania (Rao et al 1998; HASHI/ICRAF 2000), Kondoa District, Tanzania (Dejene et al 1997), and Zambia's Eastern Province (Franzel et al 2002). Kenya's Machakos District is a much-cited sustainable agriculture success story (e.g., World Bank 2007), with its finding of large productivity increases both per hectare and per capita (Boserup 1965; Tiffen et al 1994), yet follow-up work suggests these claims be treated with caution (Boyd & Slaymaker 2000; Lele & Stone 1989; Siedenburg 2006). Such impressive gains may largely reflect the significance of replenishing depleted soil organic matter, given its centrality as a determinant of the relative stability and productivity of agricultural lands (WRI 2001).

Benefits of farm trees include providing diverse tree products (e.g., wood for fuel or construction, fruit) and environmental services (e.g., soil fertilisation, windbreak). Tree costs include labour inputs, competition effects, and providing habitat to pests like ticks and the tsetse fly. Significantly, however, the financial costs of tree planting are typically marginal (Warner 1997).

*v. Technology adoption*

One fundamental characteristic of peasant agriculture is that it involves decisions by numerous households, each of which functions as an independent decision making unit (Timmer 1989), albeit one influenced by social and institutional factors. As such, farmers' NRM strategies can vary greatly from household to household, depending on their differing views of their self-interest. Farmers' decisions are central to NRM on private farms, commons areas and neighbouring open-access lands. Yet the present paper focuses on the case of private farms, since these allow the full range of alternative NRM strategies[12] including 'investing' in NR stocks, while also facilitating the isolation of local knowledge as discussed in *section xi*.

---

[8] Soil and water conserving practices common in Sub-Saharan Africa include mulching, manuring, crop rotation, intercropping, contour bunds and terracing.
[9] Agroforestry involves integrating trees into farm systems to capture ecologically and economically symbiotic interactions between system components.
[10] Integrated pest management (IPM) involves harnessing natural symbioses and predator/prey relationships as a means of managing key agricultural pests.
[11] Examples of cases where the by-products of one component become inputs to another include crop residues or tree leaves serving as livestock fodder and manure or leaf litter serving as fertilisers.

Despite their promise, adoption of regenerative technologies by peasant farmers tends to be limited (Barrett 2002). Sanchez (2002:xv) suggests that while integrated NRM holds great hope for peasant communities in Sub-Saharan Africa, the pressing challenge is to scale up what are too often isolated successes. The Millennium Project Report (2005) argues that effective technologies to address poverty "are known", but "what is needed is to apply them at scale". While findings of widespread technological innovation by peasant farmers (e.g., Reij & Waters-Bayer 2001) appear to contradict this finding of limited adoption, upon closer examination they do not. This follows because the innovations in question tend to involve relatively minor adaptations of current practice, notably experimentation with new crop varieties or planting arrangements. Bolder innovations such as experimenting with agroforestry are comparatively rare (Nielsen 2001).

Summarising, emerging NR scarcity implies a basic shift in the balance between supply of and demands upon local NRs, with growing demands beginning to outstrip resource supply. Some farmers adapt to this change by adopting regenerative practices, which foster the regeneration of key NRs. Others, however, continue using established practices that may – given current circumstances – effectively 'mine' NR stocks and undermine farm productivity. Figure 1 provides a schematic illustration of this phenomenon.

**Figure 1 Schematic of adaptation challenge facing peasant farmers lacking regular access to Green Revolution inputs**

Of course, some studies have reported spontaneous, rapid diffusion of advantageous innovative technologies. Yet these examples have generally involved marketable, high-status (i.e., 'sexy') products with short payback periods. Examples include hybrid maize in northern Nigeria (Smith et al 1994), hybrid wheat in Palanpur, India (Lanjouw & Stern 1993), coffee and tea in Kenya (Bevan et al 1989), cocoa in Sulawezi, Indonesia (Pomp & Burger 1995), and tobacco in Malawi (Place & Otsuka 2001). By contrast, regenerative technologies tend to provide mundane agricultural support services (e.g., fertilising soils, improved microclimate) and to involve longer payback periods. These cases are therefore distinct, with the latter imposing larger demands on farmers' knowledge, due to its relative subtlety.

Such distinctions notwithstanding, if both sets of technologies – low-status regenerative technologies and high-status marketable technologies – are thought to present advantageous opportunities, how might this discrepancy in farmer adoption be explained? Why might regenerative technologies sometimes be neglected, despite addressing pressing needs? If degradation of farmland harms production, why would it not be routinely addressed by farmers? Why might farmers allow key farm-based NRs such as soils and biomass stocks to be 'mined'?

*vi. Explaining halting, limited adoption of regenerative technologies*

Possible explanations for this phenomenon fall into two broad categories.

The dominant view is that peasant farmers lack incentives to redress resource degradation or adopt regenerative technologies such as agroforestry practices. Either these technologies are seen as unattractive due to hidden costs and/or inadequate benefits, or technology adoption is prevented by institutional factors such as insecure land tenure or low output prices (e.g., Gebremedhin et al 2003). Similarly, when peasant farmers act as agents of degradation, they are seen as 'forced' to degrade by sheer poverty or perverse institutions (e.g., Barrett et al 2002). In such cases, farmers' management decisions are seen as privately optimal given existing resource and institutional constraints, whether they result in technology adoption or resource degradation. Thus, Antle et al (2006) suggest that persistent land degradation despite the availability of seemingly effective restorative technologies reflects areas where such innovations are not economically viable and "a rational farmer would not invest in restoring lost soil productivity".

---

[12] NRM strategies can involve either (i) depletion of farm-based NRs via overexploitation, (ii) limited exploitation of NRs to allow their persistence and/or natural regeneration, or (iii) assisted regeneration of – i.e., 'investment' in – farm-based NRs, e.g., growing green manure or planting valued trees. Options (ii) and (iii) are both potentially optimal in the private property case, since farmers are assured access to the long-term benefit stream from these resources, while such assurance is uncertain in the case of commons areas and non-existent in the case of open access lands (Pearce 1994:3).

An alternative possible explanation for continued degradation and limited adoption of regenerative technologies is overlooked by theory but explored by the present paper and the study on which it is based, following from the Senegal anecdote. Namely, accessible regenerative technologies may be neglected by farmers, despite responding effectively to the twin challenges of resource degradation and increased demand on rural lands. Specifically, the study hypothesised that farmers' knowledge may fail to keep pace with a changing rural context, notably emerging NR scarcity[13], leading to missed opportunities and lower productivity.

Elaborating, it was hypothesised that during periods of emerging environmental scarcity, fundamentally divergent NRM patterns are pursued by households with similar assets due to differences in the LK informing NRM. Specifically, some households actively facilitate the regeneration of key NRs and integration of farm components, whilst others continue to rely on natural regeneration and show little interest in integrating farm components. Moreover, because NR stocks may be the limiting asset in areas experiencing emerging resource scarcity, these differing NRM strategies give real differences in household welfare, since some respond better to emerging challenges than others. The paper presents evidence examining this hypothesis for the case of Tanzania's Sukuma people and their management of trees.

Although the NRM strategies of peasant households are informed by their LK, optimal practice depends ultimately on the various incentives households face and how these shift over time. LK is widely believed to map these shifting incentives (see *section vii*). Where this is accurate, LK can be safely ignored as a determinant of management strategy or farm outcomes, since it will have no effect independently of 'hard' resource constraint variables such as land, labour and capital. If ever it failed to map shifting incentives, however, LK could become a key determinant of NRM outcomes in its own right, via either highlighting or failing to highlight emerging management opportunities.

**Figure 2** **Shifting incentives affecting NRM:  How well does knowledge track them?**

*vii. Knowledge in the development literature*

The 1998-99 World Development Report, *Knowledge for Development*, outlines the dominant view of knowledge and its role in development, which focuses on 'global knowledge' (GK) sanctioned by the scientific canon. While LK is recognised, it is seen as needing to be supplemented by GK for development goals to be achieved. The Report suggests that poor countries – and poor people – differ from rich ones not only because they have less capital, but also because they "have less knowledge", notably technical knowledge (World Bank 1998). Notably, GK obtained via rural education and extension contact is seen as a key determinant of technology adoption by peasant farmers (Abdulai & Binder 2006).

As an illustration, the average per capita incomes of Ghana and South Korea are compared. These figures were similar in 1960, but by the 1990s the figure for South Korea was over six times higher than that for Ghana. The Report suggests that half of this difference is due to Korea's greater success at acquiring and utilising GK. It continues that insufficient GK especially plagues the poor, and often lies at the root of their difficulties. Conversely, it suggests that increasing the poor's access to GK promises more efficient and productive agriculture, and could go far to spurring economic growth and eliminating poverty (World Bank 1998). The 2008 World Development Report, *Agriculture for Development,* maintains this view, stressing the need for rural education to enable farmers to pursue 'the new agriculture' (World Bank 2007).

LK is seen as similarly positive and unproblematic in the development literature, excepting the occasional call for avoiding an uncritical approach to LK (Oudwater & Martin 2003). Specifically, LK is characterized as sound, dynamic and well-adapted to local conditions. LK "is the basis for local-level decision making… and the main asset [of the poor] to invest in the struggle for survival. It is developed and adapted continuously to gradually changing environments" (World Bank 2008b). Within development, it is often used to tailor institutional interventions to the local context via participatory approaches based on the idea that locals know best about their circumstances and needs, though the development industry's sensitivity to LK has also been questioned (Pottier et al 2003). LK is also sometimes used to inform scientific studies (e.g., Asfaw & Agren 2007). LK of NRs and their management is thought to be particularly astute, given that rural people live and work in close proximity with these resources.

---

[13] This refers to situations where NRs have been historically abundant but have become relatively scarce over recent years.

Assumptions of LK's rigour are based partly on a priori reasoning. For instance, Titiola & Marsden (1995) argue that traditional techniques of farming and NRM represent 'best practice' under conditions that are often marginal, having been employed, tested and refined over generations. Similarly, Sillitoe (1998) argues that traditional practices based on LK successfully and sustainably exploited the resource base for centuries. This reasoning is backed up by case studies highlighting the subtlety of LK, such as the fact that !Ko women can identify 206 out of the 211 plant species growing locally (Neinz & Maguire 1979), or a study documenting the great diversity of soil knowledge within traditional societies (Alves 2005).

In sum, knowledge – whether GK or LK – is typically seen as good, with more being better and conflicts between different knowledge types or schools of thought de-emphasised. Education and extension convey GK to peasant farmers, while participatory methods convey LK to scientists and policy makers, all of which is seen as strictly advantageous, though the idea of functionally commensurate knowledge systems is also sometimes questioned (e.g., Arce & Fisher 2003).

**Table 2   Local versus global knowledge in the development literature**

Clearly, the LK that informs NRM by peasant farmers can be both powerful and subtle. Yet the present paper asks whether it may also be problematic at times, even among rational, optimising farmers. If this occurred, NRM by farmers could sometimes prove suboptimal, even where farmers *seek* to manage these resources optimally. To be clear, the present analysis *does not* hearken back to the 'bad old days' of colonialism, in which peasant farmers were sometimes portrayed as 'lazy', 'perverse' or 'irrational'. The present analysis abhors such views, and takes as its point of departure the self-evident truth that farmers are rational and optimising.

*viii. Parallels in the climate change and ecosystem services literatures*

A growing literature explores the linkages between climate change and sustainable development, notably minimising adverse impacts and maximising potential symbioses. One focus is the phenomenon of 'no regrets' technologies, which mitigate climate change yet are also privately advantageous to economic actors (i.e., have negative net cost). This potential looms large in the agriculture and forestry sectors. Yet this large potential is coupled with limited adoption. This fact is explained by diverse 'barriers' to adoption, notably market or institutional failures such as lack of information, distorted price signals, ill-defined property rights, and limited financial markets (IPCC 2007b:138-9,499,544).

Although lack of information is cited as a key barrier, this is interpreted as involving gaps in or limited access to GK. The only references to LK in the massive 2007 IPCC report involve calls to better incorporate it into GK (ibid 733). The report nonetheless touches on potential knowledge problems affecting economic actors experiencing dramatic change. Observations include:

- Extensive evidence from psychological research indicates that individual and social perceptions of risk, opinions and values influence judgement and decision-making concerning responses to climate change (IPCC 2007a:735).

- Individuals prioritise risks they face, "focusing on those they consider – rightly or wrongly – to be most significant", while cases where agents lack experience of a type of challenge "may inhibit adequate responses" (ibid 735).

- Perceptions of vulnerability and capacity to act vary among individuals and groups, and can impede adaptation to climate change. Policy needs to address such constraints by helping foster individual empowerment and action (ibid 736).

Similarly, the Millennium Ecosystem Report asserts that NR degradation is a key cause of poverty, social conflict and growing inequities, notably in rural areas where livelihoods rely directly on NRs. Yet innovative NRM provides cost-effective opportunities to address diverse development and environmental goals in a synergistic manner. Realising this prospect depends on overcoming diverse 'barriers', including gaps in GK regarding ecosystem services and their management and "poor use of existing knowledge". Besides increased study of ecosystem services and better education, other priorities for achieving sustainable management of NRs include better incorporation of "traditional and practitioners' knowledge" into GK and finding means to internalise non-marketed ecosystem values into NRM and investment decisions (MEA 2005:10-37,161).

*ix. The study context: Shinyanga Rural District, Tanzania*

The household survey was carried out in Shinyanga Rural District, Tanzania in 2000-01. This district falls in the Shinyanga Region, one of the twenty regions that make up modern-day Tanzania. Both administrative areas fall in the wider area known as 'Sukumaland', after the dominant Sukuma ethnic group. The study area is semi-arid (600-800 mm/year) and remote, an example of a 'low potential' agricultural area. Residents are overwhelmingly agro-pastoralist peasant farmers engaged in semi-subsistence production. Maize is the staple crop and cotton is the main cash crop, with other widely grown crops including sorghum, cassava, rice and sweet potato (Collinson 1972, GoT 2000).

Detailing the history of land use in Sukumaland is beyond the scope of the present paper, yet several cursory observations are needed to situate the present discussion.

Early visitors described Sukumaland as a land of plenty. Based on a visit in 1875, Henry Morgan Stanley (1899:105) wrote of "the rich country of Usukuma, where the traveller (…) need never fear starvation. The products of the rich upland were here laid at our feet, (…) plenteous stores of grain, beans, potatoes, vetches, sesamum, millet, vegetables such as melons and various garden herbs, honey and tobacco (…). The number of chickens and goats that were slaughtered by the people was enormous". Other authors report that until the early twentieth century, Shinyanga Region was extensively forested with miombo woodland species, notably *Acacia, Albizia, Brachystegia, Commiphora* and *Dalbergia* (Bradstrom 1978; Malcolm 1953; Otsyina 1992). Historically, the area is believed to have existed in a state of dynamic equilibrium between cultivated farms and grazed grasslands on the one hand and adjoining woodlands populated by wild game and tsetse flies on the other (Meertens et al 2000).

This history of natural abundance in a dynamic equilibrium shifted over the course of the twentieth century. Given the intimate association of the tsetse fly with natural vegetation (Yeoman 1967), the British colonial administration mounted an intensive campaign to remove trees and shrubs from vast areas of unpopulated woodlands as a means to control tsetse, with the approval of local people. Coupled with intensive grazing and cultivation by peasant farmers and gradual population growth, these efforts fundamentally transformed the local vegetation. Native perennial grasses and scattered shrubs were replaced by annual grasses, while native woodlands were sharply reduced (Tomecko & Tomecko 1976:13; Milne 1947:248). The rapid degradation of Shinyanga's remaining forested areas continues due to use pressures from growing human and livestock populations (GoT 2000), though local-level tree planting efforts are also observed.

This massive removal of trees has had various adverse consequences for local lands and the wider environment, including soil erosion, shortages of fuel and fodder, and water scarcity (GoT 2000). Otsyina (1995) reports acute impacts of deforestation in the area, including women having to walk up to 15 kilometres to find fuelwood and using crop residues and cattle dung as fuels instead of as natural fertilisers. Similarly, households surveyed by the author reported 'tired' soils and scarcity of fuelwood, fodder and water. Simply put, deforestation in the region has imposed major added labour demands on households while using up resources that could otherwise provide needed soil amendments to fields (Ngazi et al 1997).

The net effect is that in bad years, this is a food aid area. According to government statistics from 2000, the population of Shinyanga District had only met its carbohydrate requirements in two of the past seven years and had missed its protein requirements in all seven of these years (GoT 2000:43). A second effect is the scarcity of diverse essential products, as documented by the HASHI project.

Green Revolution inputs such as chemical fertilisers and pesticides first became available to Sukuma farmers after independence in 1961, thanks to a government programme fostering their use. Use levels slowly increased until 1986, when this trend reversed due to price rises linked to currency devaluation. From 1991 onwards, subsidies on these inputs were removed, further increasing prices and reducing use rates (Meertens & Ndege 1993). In the nineties, the removal of agricultural subsidies meant that the prices of these inputs rose rapidly, roughly trebling over their former, subsidised levels (World Bank 2000; survey households). As a result, use levels are now minimal (see Table 3). While applying manure to soils is common, quantities available tend to be insufficient to maintain soil quality (survey households; HASHI/ICRAF 2000).

**Table 3   Estimates of households using manure and mineral fertilisers in Sukumaland**

*x. Agroforestry and the Shinyanga Region*

Agroforestry practices represent an alternative management response to these emerging challenges. Historically, farmers in the area have not emphasised farm trees as a land management tool. Indeed, the Sukuma are known in Tanzania for viewing trees as a threat to their cattle, following the historical association of trees with tsetse fly in Shinyanga Region. They have a reputation for clearing large areas of forest to create virtually treeless landscapes[14] (Mshuda 1991). Yet given contemporary shortages of NRs such as fuelwood, fodder and fertile soil, the significance of farm trees has shifted fundamentally, as have the incentives for farmers to retain or cultivate them.

Diverse farm trials conducted by the HASHI/ICRAF agroforestry project[15] have demonstrated the potential of agroforestry practices to address key challenges facing households in the region. One example is rotational woodlots using *Acacia* and *Leucaena* species, which are designed to improve soil fertility and provide wood while allowing maize cropping to continue. Farm trials gave the following results: (i) trees scattered in fields reduced maize yields considerably during the first three years, after which yields began to rise again, (ii) the levels of grass fodder produced with and without trees was the same, (iii) tree biomass provided 15.4 tons of livestock fodder per hectare after three years, (iv) wood yields totalled 88.9 tons per hectare after three years, (v) available soil nitrogen approximately doubled (Otsyina et al 1996; HASHI/ICRAF 2000).

Despite this promise, government statistics highlight the project's difficulties in fostering farmer adoption of agroforestry practices. Notably, the number of seedlings raised over the period 1989-1999 shows a gradual decline over time, albeit one which coincided with phasing out of physical deliveries of tree seedlings to villages. Some observers have cited 'lack of awareness' among farmers of the importance of afforestation as a key cause[16] of this decline (GoT 2000). Whether such claims are right or wrong, this question clearly demands critical scrutiny, as do the broader reasons for the limited farmer uptake of agroforestry practices, given the needs they seem to address.

The HASHI project's own findings likewise call the success of its work into question. In 1999, Shinyanga Region farmers were surveyed about their experience with agroforestry following project outreach efforts. Farmers were found to view trees primarily as sources of fuel and construction wood, while relatively few saw trees as a means of addressing pressing problems such as infertile soil and fodder scarcity (8 and 18 percent of households respecively) (ICRAF 2000).

*II. Findings*

*xi. Gathering empirical data, conducting data analysis*

The above hypothesis was examined using fresh data from an in-depth household survey of 350 randomly selected households in twenty villages conducted in Shinyanga District, Tanzania in 2000-01. The study sought to explain observed tree management patterns by examining their relation to diverse farm and household characteristics, including the LK informing each household's farm management practices. It focused on LK vis-à-vis trees, notably on farmers' recognition and valuation of economically relevant aspects of trees. These aspects include products such as fruit and wood, services such as soil fertilisation and acting as a windbreak, and costs such as shading crops and providing habitat for crop or livestock pests.

---

[14] The key exception to this rule of de-emphasising trees is the traditional *ngitiri* system of conserving designated lands for dry season pasture (Mlenge 2005).

[15] The HASHI/ICRAF agroforestry project seeks to develop appropriate technologies for farm households in Shinyanga Region, then foster their adoption. It is a collaboration between the Tanzanian government and World Agroforestry Centre, an international research institute.

[16] Another cause sometimes cited is technologies ill-suited to the local context, whether due to failing to fit into farmers' livelihood strategies or simply being ineffective. E.g., the HASHI/ICRAF 1999 Annual Report states, "Nitrogen-fixing trees have the potential to restore the fertility of farmland as well as to provide fodder and wood. This potential has not been exploited due to the lack of appropriate technologies to integrate trees into the existing land-use system".

To facilitate the identification of knowledge gaps, the study emphasised NRM on private farms. Associated 'environmental values' (goods, services, costs) are thus private values that accrue to the land owner. As such, the incentives guiding management of these resources are straightforward (i.e., "conserve or cultivate" a resource where it is valuable, "remove or deplete" it where it isn't), as opposed to the case of common property or open access resources, where 'perverse' incentives may favour resource depletion even where the resource is highly valued. Because farmers working private farms have secure access to the long-run benefit stream associated with farm-based NRs, they face clear incentives to optimise their management based on *all* relevant aspects of these resources. As such, cases where households do not recognise aspects of trees that have economic value in the contemporary context may be interpreted as knowledge gaps. The study assumed that a given aspect of a tree has economic value when it fills a clear household need and some farmers recognise this aspect as valuable in their statements during household surveys. Where others on similar farms do *not* recognise this aspect of the tree, this was interpreted as a knowledge gap. Thus, where *some* farmers recognise the soil fertilisation function of a given tree but others do not in an area where soils are poor and soil nutrients scarce, the latter group may be said to have a knowledge gap.

To be as certain as possible that elusive LK data were captured, measures were taken to ensure that farmer knowledge about trees was fully explored, and that aspects of knowledge were not overlooked based on misunderstandings. On the one hand, interviews involved open ended questions such as, "Which trees are most significant to the household's welfare?", and "What are the linkages between these trees and farm enterprises such as field cropping and livestock?" On the other hand, they included highly specific questions about the various potential goods, services and costs associated with each tree type cited as significant to the household. For each such tree, a list of possible tree characteristics was read to the farmer, asking him/her to say which of these applied in their case. At the end of this process, it was deemed that if a farmer does not express a given type of knowledge, he/she may lack this knowledge.

The analysis involved both a regression analysis of a set of tree management models and an associations analysis of species-specific tree knowledge. The former sought to assess the significance of LK to explaining observed tree management patterns, while the latter sought to discern patterns within the wider body of LK. The tree management models analysed reflect the key distinctions among households' tree management strategies. Each emphasises a different aspect of tree management and provides a different window onto management and its determinants.

*xii. Findings of regressions analysis*

Tree management by peasant households in the district is very mixed, in terms of both species and management practices (see Tables 4 and 5). Simply put, farmers follow fundamentally divergent practices. As seen below, farmers also have divergent ideas about trees, despite numerous tree planting initiatives over the years by both government and development agencies.

**Table 4** Tree species cited by at least 10% of survey households
**Table 5** Raw tree management data used in analysis

The analysis also showed that the causality of tree management patterns is complex, since considering household resources such as land, labour and capital only explained a relatively modest proportion of observed variation. Integrating 'global knowledge' variables such as schooling and contact with extensionists into regression equations increased their explanatory power somewhat. Yet integrating 'local knowledge' variables into the regression equations was much more significant, often doubling the explanatory power of tree management models. See Table 6 for variables found to be significant to the regression analysis, Table 7 for results of this analysis, and Table 8 for a breakdown of the determinants of the nine tree management models.

**Table 6** Variables found to be significant and relative frequencies
**Table 7** Explanatory power (i.e., R2) of different versions of the logit models
**Table 8** Factors found to be significant in tree management models

Thus, farmers' perceptions of trees and ideas about 'best practice' tree management appear to diverge in ways independent of their tangible resource constraints, and these differences seem to strongly impact on farmers' tree management practice. Several alternative interpretations may be distinguished.

One interpretation is that these 'LK variables' actually reflect subtle differences in resource constraints that were not accounted for in the analysis. A second is that they may reflect differences in household preferences. In both cases, the implication is that these variables may not reflect knowledge differences at all. Such explanations cannot be ruled out. Yet the former alternative was controlled for insofar as possible, while the role of subjective preferences is arguably minimised by the poverty of survey households, which means they can ill-afford to indulge whims that might interfere with securing basic needs.

The third possible interpretation is that the findings strongly support the study hypothesis. That is, the findings may be interpreted as demonstrating that LK sometimes fails to reflect the incentives faced by households, and hence that farmers' knowledge may influence their management strategies *independently* of tangible opportunities and constraints. Reiterating, it was hypothesised that this occurs only under particular circumstances, namely where peasant communities face emerging NR scarcity associated with intensive pressures on local NRs – a challenging situation for which their prior experience may prepare them poorly. These conditions hold in contemporary Shinyanga Rural District, Tanzania.

Assuming this third interpretation is correct, differences in LK could critically determine outcomes vis-à-vis tree management, and perhaps NRM more generally. For instance, if a farmer does not recognise the soil fertilisation function of a given tree, this may well prevent him using it to address a soil fertility problem, or perceiving the loss this may cause the household. Whether or not a farmer demonstrates knowledge of a given tree characteristic may thus be relevant to understanding his/her management decisions.

*xiii. Associations analysis of LK data*

To further elucidate the regression findings, the study ran a series of correlations among study variables. These sought to identify patterns within the wider body of LK, notably among species-specific tree knowledge variables. They also examined associations between these knowledge variables and diverse resource and management variables[17]. The premise was that if the study could demonstrate which aspects of LK tend to go together, or whether different types of LK form distinct knowledge profiles, this could shed light on observed knowledge differences.

Correlations can be used to ascribe trends among households, yet do not reveal causality. Given two correlated variables, it is unclear which causes which, or if their association is caused by a third variable or variables (Black 1999:215; Norusis 2002). As such, this part of the analysis was purely exploratory. The findings nonetheless seem to support the suggestion that observed management differences are partly due to fundamental differences in LK. Notably, the statistical associations summarised in Figure 3 suggest that farmers' thinking vis-à-vis trees falls into three broad categories, as does their observed farm management practice. The qualitative evidence cited in Siedenburg (2005) also supports this interpretation[18].

**Figure 3   Diagram of the associations analysis**

Category one may be characterised as disinterest in trees and their potential role within the farm system. These households emphasise low-value tree products and obvious tree services, yet neglect high-value products and subtle services.

Category two involves viewing trees as the source of high-value tree products such as fruit and construction wood. These households also stress harmful impacts of trees on crops via soil effects or shading, yet neglect their capacity to provide agricultural support services when managed appropriately.

---

[17] For each potential association, the study obtained a chi-squared statistic, a gamma statistic and spearman's rho, to assess statistical significance and obtain comparable measures of strength and directionality.
[18] These data showed that tree management practice in Shinyanga District, Tanzania falls into three broad categories – 'did not plant trees', 'planted trees then left them alone', 'planted and watered trees' – and that farmers' comments about trees fundamentally differed between these categories. The three distinct perspectives on trees conveyed by farmers in this earlier paper roughly mirror the categories derived from the associations analysis summarised in Figure 3.

Category three involves viewing trees as providing diverse products[19] and agricultural support services, notably the critical service of soil fertilisation. It represents a broad view of trees that includes their potential to harness positive feedback loops within the farm system. The tendency of these households to cultivate leguminous crops and vegetables suggests dynamism and a readiness to respond proactively to increasing pressure on the local NR base.

Figure 3 also includes interpretative comments on how these three broad knowledge categories compare with one another. Such comparisons are significant, since individual households may move between categories in response to shifting incentives or increasing knowledge. Categories 1 – 3 represent progressively higher implicit valuations of trees, as well as an ever more comprehensive view of trees and interest in trees as a farm management tool. A second important trend across the three categories is that the households in categories two and three tend to be better off than those in category one, as reflected in their tendency to have larger farms, more cattle, better familiarity with Swahili, and greater access to radios. This may be interpreted either as evidence that more comprehensive tree knowledge is partly a function of relative privilege, or that having better knowledge of trees may help farmers seize opportunities and accrue wealth.

*III. Discussion*
*xiv. Possible reasons for observed gaps in LK, given emerging resource scarcity*

Possible reasons for observed gaps in LK include questions about both demands upon LK and the supply of LK. The reasons cited relate to LK and NRM in peasant communities, but many also relate to broader issues of knowledge and optimal NRM, including how NRs are managed by institutions and in rich societies. The simple fact is that given emerging NR scarcity, NRM may prove problematic in diverse contexts for broadly similar reasons.

Emerging NR scarcity places great demands on the knowledge informing NRM (see Table 9a). 'Demand' factors include ways that the NRM problem facing farmers is complex, given emerging resource scarcity. LK must negotiate these complexities and adapt to any changes, if it is to inform optimal NRM. Most fundamentally, the fact that NRs may be regenerated either naturally or via human action raises questions about when the latter becomes advantageous. At this point, maintaining optimal NRM may require radical management innovations, such as cultivating trees to provide fuel rather than gathering wood from the bush or actively maintaining soil fertility rather than relying on fallowing.

**Table 9a   Given emerging resource scarcity, the demands upon LK informing NRM are great**

At the same time, the responsiveness of LK to change may be problematic during times of emerging NR scarcity (see Table 9b). Concerns about the readiness of LK vis-à-vis NRM to adapt to a changing rural context stem principally from two characteristics of NRs: that they were historically abundant and that they regenerate spontaneously under normal conditions. Both traits may encourage farmers and others to see environmental goods and services as 'free gifts of nature' and hence to take their presence and reproduction for granted. Such a view is unproblematic as long as NRs are abundant, but this may change where NRs become scarce if they provide needed but unheralded goods and services.

**Table 9b   Reasons for concern that LK may adapt only slowly to emerging resource scarcity**

The responsiveness of LK to change may also be constrained by cultural embedding, whereby ideas about appropriate NRM that developed at a time when NRs were abundant become rigid social norms possibly reflected in local mythology[20]. In Sukumaland, for instance, historical associations of trees with agricultural pests such as tsetse fly and quelea quelea birds could crystallise into cultural biases against trees that hamper farmers' capacity to recognise emerging opportunities involving trees.

---

[19] Although tree products are not listed as key correlates in this category, this follows because different tree products were cited by different households, but in all cases at least some high-value tree products were also cited.

[20] "If you are steeped in social norms of behaviour... you do not calculate every five minutes how you should behave.. you follow the norms (Dasgupta 1996)." Yet established practices may remain optimal only as long as the local context remains roughly constant.

Flawed information provision to rural communities may likewise constrain LK. Such information may mislead and obfuscate instead of informing and empowering. For instance, it may advocate technologies that are inaccessible to poor households or condemn useful traditional practices as backwards. The quality of LK could suffer, given the likely impact on farmers of views and strategies advocated by powerful outside actors such as extensionists. Figure 4 illustrates the problem.

**Figure 4 "A clean field is a good field"**

This series of images – as reported by Mshana (1992:212-18) – is taken from a primary school textbook lesson titled *Principles of Good Agriculture* that is still widely used in Tanzanian schools (Lugendo pers com 2001). Yet this lesson may mislead farmers, notably those with degraded farms who lack access to purchased farm inputs, since it forcefully asserts that trees should be removed from fields. The lesson begins, "So as to be successful in your farming by achieving high crop yields, you are obliged to follow the following good agricultural principles," then concludes, "Advise your parents on… stages to be followed." Worryingly, Mshana reports that most educators only "realized for the first time there was a gap between the contents of the book and the surrounding environment" when interviewed for his research, suggesting hesitancy to question received wisdom.

*xv. The farmer's management decision revisited*

If LK sometimes proves problematic for the foregoing reasons, the danger is that some peasant farmers may not recognise the point at which adopting regenerative practices becomes advantageous, and hence may fail to make the transition from relying on natural regeneration to actively facilitating regeneration. This is most likely to occur where NR scarcity emerges rapidly, since in such cases dramatic management changes may be needed and past experience may be a poor guide to current challenges. This contrasts with the case of LK vis-à-vis *static* phenomena such as ethnobotany, whose accuracy is often cited as evidence of the subtlety of LK.

The NRM challenge facing the peasant farmer is to identify the privately optimal management strategy for each key NR affecting his farm system or livelihood. Three broad NRM strategies may be distinguished, namely exploitation, cultivation, and stewardship. Each can be applied to any given NR at any given time, based on the farmers' management decision. Each of these three strategies can be optimal depending on the local abundance of the NR, its role(s) in the farm system and wider economy / ecosystem, and the availability of substitutes. Farm households can move between these strategies in response to changing conditions, yet could potentially persist with a strategy too long due to problematic LK.

**Table 10 Comparing the three broad NRM strategies available to farmers**

Clearly, the incentives facing farmers are key determinants of farmers' NRM decisions. Yet the findings reported suggest that favourable incentives may not suffice to secure farmer adoption of advantageous technologies involving innovative NRM. That is, the incentives created by shifting resource constraints and institutional arrangements may be *necessary but not sufficient* to secure an innovative NRM response that is optimal under the new conditions. While farmers surely *tend to* respond to shifting incentives, since they are rational and observe their surroundings, the evidence presented suggests that LK does not always track shifting incentives – or does so only with a significant time lag – with potentially serious welfare implications. Paradoxically, 'standing still' may be a risky strategy in a situation characterised by 'moving goalposts' of NR abundance, since farmers could miss out on advantageous NR cultivation or stewardship opportunities.

Within Shinyanga Region, a broad trend has been observed for farm households to compensate for reduced access to off-farm trees – and the various benefits they provide – by growing farm-based trees (Otsyina et al 1997). The present paper assesses the general applicability of this claim in Shinyanga District, following from the suggestion by Scherr & Hazell (1994) that farmers may not automatically adapt their management practices to emerging land scarcity.

*xvi. Theoretical models and LK may dovetail, obscuring gaps and prospects from development professionals*

Critically, the idea that LK may be problematic is ignored by peasant farmers and outsiders alike. These farmers, like people everywhere, do what they think best and do not normally doubt their perspective. Yet this possibility is also neglected by development practitioners and researchers, since it is assumed away by current development theory. Specifically, theory assumes away the possibility that advantageous, accessible opportunities may be overlooked by farmers. This was seen in the LK literature cited in section vii above, and is also clear from the agricultural economics literature cited below. These different strands of formal theory dovetail around a view of local knowledge as unproblematic. A second parallel is that both farmers and development professionals face the same broad challenge, namely adapting their ideas of key opportunities and constraints – and hence 'best practice' management – to a situation of emerging NR scarcity.

The 'knowledge problem' facing peasant communities is thus two-fold, given emerging NR scarcity. Empirical evidence reported in the present paper suggests that the LK informing NRM by peasant farmers may fail to track emerging opportunities. Yet development theory fails to track these same opportunities. Thus, farmers may miss accessible opportunities and suffer losses, without either farmers or development professionals recognising the missed opportunities.

**Table 11   The two facets of the knowledge problem:  Farmer practice, development theory**

The fact that development professionals and farmers may share the same oversight makes instances of problematic LK especially tenacious and difficult to detect. The end result is that both groups may neglect opportunities involving addressing gaps in LK, since both are plagued by the same 'blind spot'. Conversely, filling in this lacuna by identifying and addressing gaps in LK promises to help reduce poverty, bringing unanticipated welfare gains.

*xvii. The peasant farming literature and the farmer's NRM decision*

Several leading conceptual models from the agricultural economics literature are highlighted. Despite their differences, these models share certain characteristics. Most notably, they all assert farmers' responsiveness to incentives, including responses involving the adoption of innovative NRM practices. This literature thus masks potential problems with either LK or NRM by peasant farmers.

*a. The 'poor but efficient peasant' thesis*

Since Schultz's landmark book of 1964, it has been argued that peasant farmers combine the resources available to them optimally in the aim of securing their basic needs. Simply put, peasant farmers are 'poor but efficient'. The implication is that 'no appreciable increase in agricultural production is to be had by reallocating the factors at the disposal of farmers' (Schultz 1964:39). This thesis has been highly influential. Ali & Byerlee (1991) call it one of the most enduring themes in development economics, while Ellis & Biggs (2001) suggest that 'agricultural growth based on small-farm efficiency' is the paradigm that has most dominated rural development thinking over the last half-century.

Adherence to the 'poor but efficient peasant' thesis has been fuelled by both a priori logic and empirical studies. The former runs thus: peasant farmers have long depended on local NRs, so it makes sense that they would be intimately familiar with them and manage them efficiently. Empirical studies of farmer efficiency have found that farmers respond to price movements and that observed ratios of marginal value product to marginal factor price suggest efficiency, yet these measures are controversial[21] (Ali & Byerlee 1991).

One useful impact of the 'poor but efficient peasant' thesis has been to strongly assert the idea of farmer rationality (Hoff et al 1993). The trouble is that this thesis obscures the distinction between seeking to be efficient and achieving it.

*b. The Boserup hypothesis*

Using Machakos District, Kenya as a case study, the Boserup hypothesis suggests that population growth coupled with market access can lead to large productivity increases while also reversing trends towards resource degradation. The essence of the model is that rural people respond positively to the challenge posed by higher population densities through appropriate investments, with the result that extensive fallow systems change spontaneously into intensive systems. A key aspect of this positive response is farmer 'investments' in land quality via adopting innovative NRM practices such as tree planting and terracing (Boserup 1965; Tiffin et al 1994). While information about competing technologies is seen as necessary for adaptation to change, this information is believed to follow naturally from increased population density and market activity.

---

[21] These studies found that ratios of marginal value product to marginal factor price were 'sufficiently close to one' to confirm that peasants are allocatively efficient. Yet these studies were highly conservative about rejecting the null hypothesis that k was equal to one. Thus, k values as high as 3.6 and as low as 0.6 were interpreted as 'sufficiently close to one'. Other limits to these studies included their focus on farmers employing Green Revolution technologies and their dependence on how production functions were specified, e.g., which types of labour use were factored into equations (Ali & Byerlee 1991).

*c. The induced innovation thesis*

Induced innovation theory suggests that farmers innovate in response to changes in local opportunities and constraints (Binswanger & Ruttan 1978; Koppel 1995). Simply put, it suggests that price changes spur adaptation to change and invention. While this model has often been associated with Green Revolution technologies and commercial agriculture (e.g., Koppel & Oasa 1987), Binswanger & Ruttan (1978) cite its relevance to the non-market situations commonly encountered in peasant agriculture, where the opportunity costs of non-marketed resources play the role of market prices as indicators of value. In such situations, innovative NRM by farmers is expected – such as investment in agroforestry practices or conservation measures – in response to incentives associated with population growth and market pressures (Barbier & Hazell 1998; Scherr 1997). Critically, the model sees innovation as driven by a dialectical interaction between technical and institutional change, including enabling interventions such as farmer education and market-friendly policies (Hayami & Ruttan 1985).

*d. The 'tragedy of the commons' model*

Although it has been subjected to critiques (e.g., Pearce & Warford 1993:239) and counterexamples (e.g., Feeny et al 1990), the "tragedy of the commons" remains a potent thesis within the development literature. In a seminal paper, Hardin (1968) argues that social or institutional factors can create perverse incentives that prevent farming communities from successfully resolving the NRM problem. The argument is illustrated using the case of communal grazing, but it applies equally to other NRs that are obtained from communal or off-farm areas, such as fuelwood or wildlife.

The problem is that the costs of a marginal increase in a household's herd are shared among all the users of the commons while the benefits accrue only to the household. Given this situation, "the rational herdsman concludes that the only sensible course for him to pursue is to add another animal to his herd". Such management arrangements are unproblematic as long as the NR in question remains abundant, yet they become problematic in situations of resource scarcity. When population density surpasses a certain level, "the inherent logic of the commons remorselessly generates tragedy". The tragedy is that each household seeks to increase its herd without limit in a world that it limited.

To address this problem, Hardin calls for either privatisation of communal resources or government-imposed laws or taxes that simulate private incentives. "The tragedy of the commons… is averted by private property, or something formally like it… [since] private property… deters us from exhausting the positive resources of the earth". This solution suggests that the NRM problem is wholly due to perverse incentives, and hence will sort itself out once these are addressed.

*xviii. Impacts of this theory on peasant agriculture*

Despite growing concerns that resource degradation is undermining agricultural production in poor areas (e.g., World Bank 2007), the role of NRs in underpinning production has often been neglected in the development economics literature (Dasgupta 2001). Similarly, innovative NRM as a potential solution has arguably been less of a focus than it merits, notably in areas where access to Green Revolution inputs is limited. This neglect and de-emphasis can be traced in part to the theoretical expectation that rural people spontaneously seize advantageous NRM opportunities based on sound LK, obviating the need for explicit policy focus in this area. Environmental considerations may also be de-emphasised in participatory consultations on rural development insofar as farmers do not perceive problems with NRM, as in the discussions conducted for Tanzania's Poverty Reduction Strategy Program (PRSP 2000).

Despite this underlying logic within the theoretical canon, some authors nonetheless stress the importance of securing adoption of innovative NRM practices by peasant farmers, based on a sense that this is not occurring sufficiently at present. For instance, Barrett et al (2002) cite farmer adoption of such practices as the great pressing challenge facing African agriculture. Yet the incomplete theoretical basis for such work – the lack of answers to the question, "Why is NRM by farmers problematic then?" – hampers its efforts to identify and address constraints to optimal NRM. Moreover, in the absence of a theoretical construct which allows LK of NRM to be problematic, researchers confronted with resource degradation may tie themselves into improbable knots. For instance, in writing about resource degradation in Kondoa District, Tanzania, Dejene et al (1997) first cite 'poor' NRM practices by farmers, then pay homage to the ingenuity and resourcefulness of local people.

*xix. Recognition that LK may be problematic highlights new opportunities*

The present paper highlights opportunities masked by both LK and existing theory informing work on peasant production. It questions the assumptions about LK made in the development literature and proposes an alternative perspective on the knowledge informing NRM in peasant communities facing emerging NR scarcity. Once this knowledge is seen as potentially flawed, new opportunities come into view, notably opportunities to profitably pursue sustainable agriculture.

While sustainable agriculture has long been recognised as a potentially promising means of addressing the challenges facing poor farmers, the viability of specific technologies has been assessed based largely on the adoption response of target beneficiaries, i.e., on farmers' 'revealed preferences'. Farmer assessments are seen as particularly important for agroforestry practices due to their complexity, notably their diverse costs and benefits and status as multi-year investments. Farmers are thought to be well-placed to assess the profitability of these practices under farm conditions, based on the idea that they allocate their limited resources among competing enterprises in ways determined by their bio-physical and socio-economic constraints and household priorities (Franzel & Scherr 2002).

Although this logic is compelling, its practical effect has been to lead development professionals to conclude that agroforestry technologies are often ineffective at addressing the challenges farmers face, despite the promise shown in field trials. This conclusion follows from the observation that farmers' adoption of these technologies has often been hesitant and limited. If, however, farmers sometimes neglect key aspects of agroforestry technologies, then addressing these knowledge gaps could reveal to farmers ways in which these technologies represent untapped opportunities for livelihood gains.

Insofar as opportunities are constrained by knowledge gaps, interventions to harness them will need to emphasise information provision. Specifically, targeted information provision may be needed to address identified knowledge gaps, in order to ensure that farmers can weigh up lucidly the alternative management options before them. This fits with the existing call for rural education to be reoriented towards problem solving and entrepreneurship in order to overcome "commonplace psychological and cultural barriers to innovation" (IFPRI 2007:16).

Table 12 could provide the basis for a simple information provision module to address the gaps highlighted in this paper. The table lists key needs of peasant households, then cites alternative ways of meeting each need. These different answers may be grouped into three broad agricultural strategies, (i) extensive agriculture, or 'shifting cultivation', (ii) Green Revolution agriculture, (iii) intensive sustainable agriculture. The module could seek to ensure that farmers are familiar with the main alternative means of meeting key household needs. Farmers could discuss the alternative means of meeting each need in the local context, being sure to mention options from each of the three broad agricultural strategies in each case.

<u>**Table 12**</u>  **Alternative means for peasant farmers to meet key household needs**

*xx. Implications of the study findings*

At its simplest, the present paper's conclusion is that peasant households either do or do not actively foster the regeneration of key farm-based NRs when these become degraded, and that differences in LK may be a key determinant of this decision.

Several theoretical and policy implications may be discerned. Specifically, the paper…
a. Elucidates a neglected determinant of whether the linkage between rural livelihoods and local environments forms a 'vicious circle' of stagnation and degradation or a 'virtuous circle' of

      dynamism and resilience, which may vary from community to community and even farm to farm.

b. Helps explain the ongoing NR 'mining' on many peasant farms and the limited adoption of sustainable agriculture technologies by peasant farmers generally.

c. Suggests concrete measures to help farmers seize advantageous yet neglected NRM opportunities, namely tailoring information provision via extensionists, schooling and radio to address identified knowledge gaps. Such information could empower farmers, providing a pathway out of poverty by securing the diffusion of 'best practice' NRM.

d. Beyond its direct relevance to poverty reduction, such tailored information provision could also prove a key complement to any future efforts to engage peasant farmers in ecosystem service delivery, for instance by integrating them into the global carbon market[22].

Summarizing the paper's implications, significant opportunities for improving the welfare of peasant households are masked by the blanket assumption that LK is well adapted to the local context, and hence that the NRM it informs is roughly optimal given the constraints households face. Yet recognizing that LK may be problematic under certain conditions brings these opportunities into view, while also suggesting steps that could help harness them.

*xxi. Conclusions*

      Peasant farmers in the contemporary developing world often find themselves in a difficult situation. Notably, poverty, stagnant agricultural production and environmental degradation are pressing, grave problems, particularly in Africa. Meeting the poverty reduction challenge will involve defining the investments needed to address constraints on key target groups. Thus, the 2005 Millennium Project Report asserts that "both villages and cities can be empowered to become part of global economic growth *if they are empowered with the infrastructure and knowledge to do so*" (author's emphasis).

      Based on fresh empirical data, the paper examines whether gaps in LK vis-à-vis NRM may be detected, and whether addressing such gaps may represent a promising investment for reducing poverty in peasant communities. Instead of speaking loosely of the need of such communities for 'literacy, numeracy and marketable skills' (Millennium Project Report 2005:13), the paper seeks to identify critical gaps in LK as a basis for tailored information provision.

      Constructive criticism of farmers' knowledge or management practices in the literature is rare (e.g., Lipton 1968), due in part to this debate being framed as a choice between seeing peasants as either 'efficient' or 'irrational' (Adams 1986). Simply put, would-be dissenters may fear being branded as neo-colonial. Clearly, we must avoid any hint of hearkening back to the 'bad old days' of colonialism or to harsh views of peasant farmers that reflect either the observer's ignorance or racist condescension. Yet even a worthy idea can become a conceptual straightjacket, hampering analysis of potentially vital issues.

      Practically speaking, the fact that the Shinyanga study identifies gaps in local knowledge vis-à-vis trees is great news, since it offers hope that hard-pressed farmers could powerfully 'help themselves' if catalysed by tailored information provision. This remedy need not imply a major new developmentintervention, and could simply involve revising the information content delivered by existing information provision channels such as schools, extension services and radio.

      Addressing these knowledge gaps not only holds out hope of improved material welfare, but could also empower farmers, helping restore their agency by highlighting ways for them to ameliorate their situation. This contrasts with the literature on 'poverty traps' or references to farmers being 'forced' to degrade vital NRs, whereby the poor are seen as unable to make needed investments or

---

[22] Recent work in this area by Lal (2007), the BioCarbon Fund and others holds out the exciting prospect that this market could in time serve as a means to convert peasant lands into a vast carbon sink by paying farmers for provision of carbon sequestration services, simultaneously mitigating climate change and reducing poverty. Incorporating these areas into the carbon market could create powerful new incentives for farmers to adopt regenerative technologies, which tend to sequester carbon in both biomass and soil. Yet getting these new incentives to 'bite' – i.e., to deliver a widespread adoption response by farmers – could depend in part on addressing gaps in LK, to ensure that farmers fully appreciate the different aspects of these opportunities.

avoid harmful actions.  This empowerment could potentially be powerfully enhanced if such information provision were combined with integration of peasant communities into the emerging environmental services markets, for instance as providers of carbon sequestration services to the carbon market.  In this case, farmers would know that they were delivering valued services to the global market and being compensated accordingly, as opposed to seeing themselves as perennial supplicants.

Given the gravity and complexity of the problems facing many peasant farmers, development professionals must be sure to help farmers to meet challenges, not impede them.  One potential pitfall lies in adhering to flawed theoretical constructs.  At their best, these constructs foster accurate analysis and effective intervention.  Yet this requires that they be accurate shorthand approximations of the realities they represent.  To ensure this, experts and theoreticians must continually reassess their ideas and assumptions, especially when target communities face rapid change.

**Table 1**   Peasant farming and competing pathways to agricultural innovation

|  | **Green Revolution Model** | **Sustainable Agriculture Model** |
|---|---|---|
| **Objectives** | Agricultural growth via market participation | Bolstering subsistence production OR providing a firm foundation for growth and market participation, including organic food |
| **Key inputs into farming system** | 'Purchased inputs' such as hybrid seeds and chemical fertilisers and pesticides | 'Natural inputs' drawn from local NRs, e.g., mulch from crop residues or leaf litter, manure, tree fodder, microclimate services |
| **Role of local NRs** | Relatively unimportant | Provide key production inputs |
| **Means of accessing inputs** | Purchased from markets, perhaps with help from state or donors | Differential allocation of household labour to regenerate/maintain key natural resource stocks, i.e., farmer's management decision |

**Figure 1**   Schematic of adapation challenge facing peasant farmers lacking regular access to Green Revolution inputs

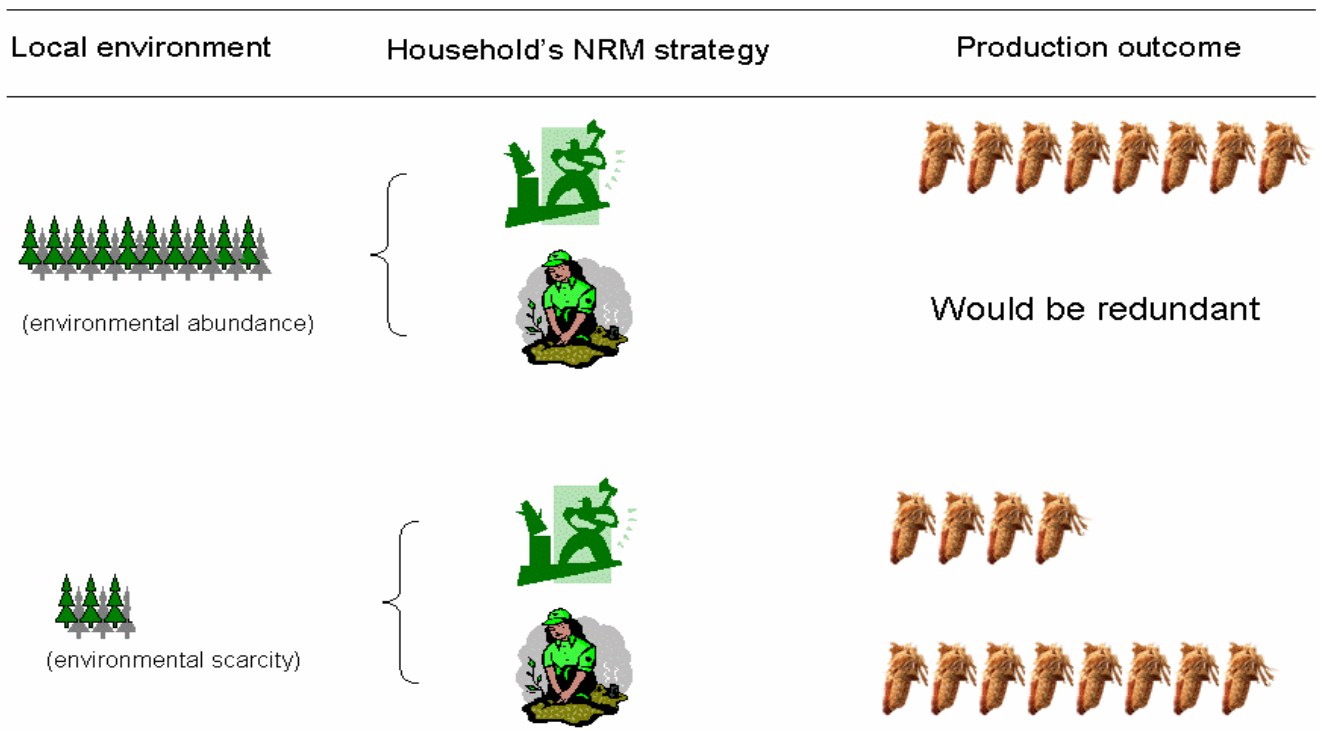

**Figure 2   Shifting incentives affecting NRM:  How well does knowledge track them?**

⇒ The NRM strategies of peasant households depend on incentives faced *and* on how LK maps shifts in incentives
⇒ For NRM to remain optimal, the LK informing it must map these shifts

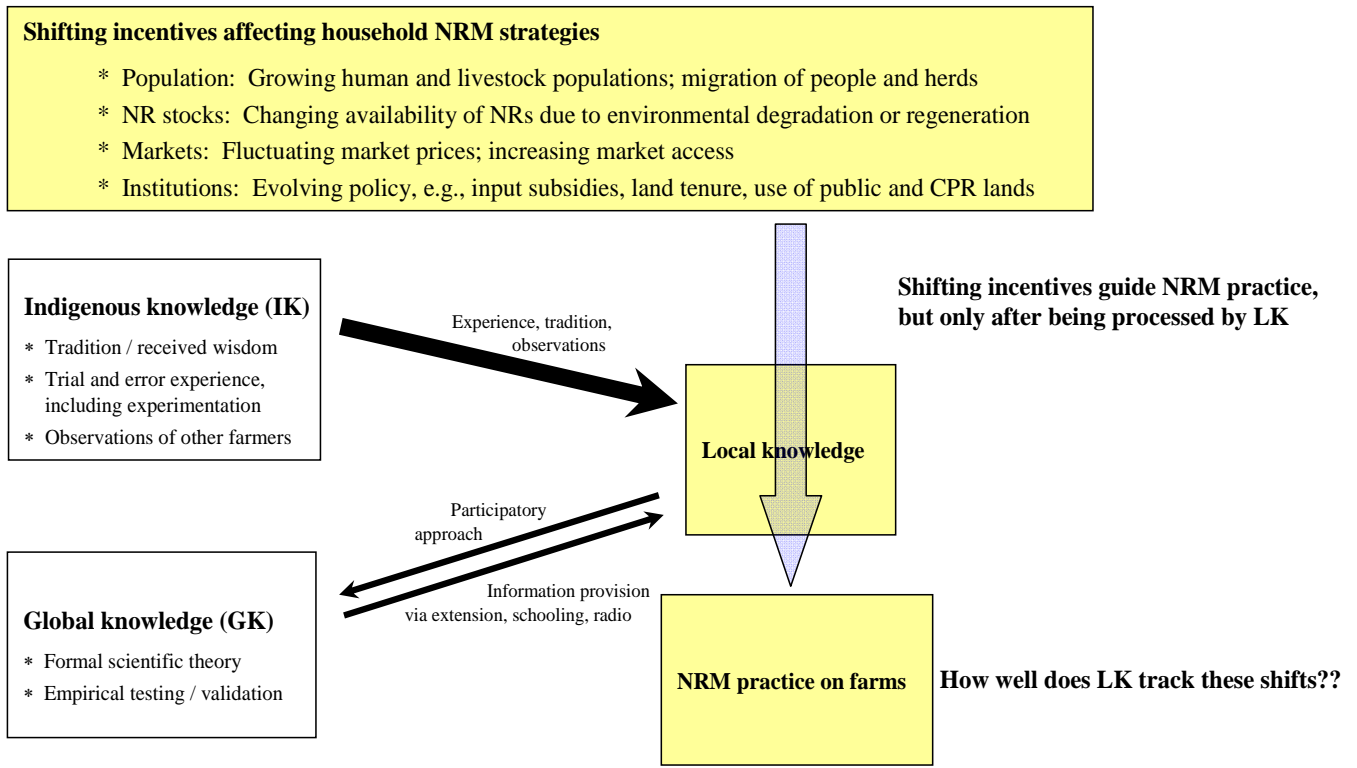

**Table 2   Local versus global knowledge in the development literature**

|  | Local knowledge | Global knowledge |
|---|---|---|
| **Applicability** | Often seen as specific to the local socio-cultural, agro-ecological context | Meant to have universal applicability |
| **Uses** | Guides the decisions of peasant farmers, informs interventions via participatory approach | Key determinant of development, notably via schooling, extension services |
| **Validation** | Wealth of practical experience based on day-to-day management and observation | Scientific method, technological applications |
| **Transmission** | Transmitted orally from fellow farmers or derived from own observations and experience | Transmitted via formal information provision media such as education, extension or radio |
| **Status** | May be low | Tends to be high |
| **Costs** | Relatively low, involving observing, listening, experimenting w/ competing practices | May be inaccessible to poor farmers due to costs associated with schooling and radios, or to poor quality rural education |
| **Knowledge gaps** | Gaps vis-a-vis local NRs not foreseen, with LK viewed as dynamic and well-suited to the local context | Gaps foreseen, with GK secured only where it is accessible, affordable and cost-effective |

**Table 3   Estimates of households using manure and mineral fertilisers in Sukumaland**

| Year | 1880 | 1945 | 1955 | 1962 | 1990-91 | 2000-01 |
|---|---|---|---|---|---|---|
| **Source** | Tomecko & Tomecko (1976) | Rounce (1949) | McLoughlin (1967) | Collinson (1963) | Ahmed et al (1990); Ebong et al (1991); Bantje (1991); Meerten & Ndege (1993) | Siedenburg (2003) |
| **Manuring** | 50% | 3-5% | 5-10% | 3% | 5-50% | 45% |
| **Chemical fertilisers** | - | - | - | - | 3-30% | 1% |

### Table 4　Tree species cited by at least 10% of survey households

| Botanical name (**exotic fruit tree) (*indigenous fruit tree) | Common name (English) | Common name (Swahili) | Households citing tree | % citing it as most significant tree | % citing it among 3 most significant trees |
|---|---|---|---|---|---|
| Acacia nilotica | Prickly acacia | *Mihale* | 122 | 21 | 35 |
| Acacia tortilis | Umbrella thorn | *Mgunga* | 101 | 13 | 29 |
| Mangifera indica** | Mango | *Maembe* | 61 | 10 | 17 |
| Tamarindus indica* | Tamarind | *Mkwaju* | 55 | 4 | 16 |
| Cassia siamea** | Sweet orange | *Mchongoma* | 51 | 5 | 14 |
| Acacia drepanolobium | - | *Malula* | 49 | 2 | 14 |
| Acacia polyacantha | White thorn | *Nguu* | 47 | 2 | 14 |
| Balanites aegyptica* | Desert date | *Miyuguyu* | 39 | 2 | 11 |
| Azadiracta indica | Neem | *Mwarubaini* | 36 | 4 | 10 |
| Albizia harveyi | Rain tree | *Mpogolo* | 35 | 5 | 10 |
| Leucaena leucocephala | Lusina | *Lusina* | 35 | 4 | 11 |

Source: Shinyanga District household survey, 2000

### Table 5　Raw tree management data used in analysis

| Variable | Brief description | Categories | Frequencies | Cases |
|---|---|---|---|---|
| **Tree numbers** | Total number of trees on the farm | 5 or less<br>6 – 20<br>21 or more | 30%<br>34<br>36 | 340 |
| **Tree management** | Cultivation intensity | Systematically removed from farm<br>Sometimes retained<br>Simply planted<br>Planted and watered | 18<br>49<br>23<br>10 | 335 |
| **Tree types** | Broad type of trees retained or cultivated | Exotic fruit tree<br>Indigenous fruit tree<br>Non-fruit tree<br>None cultivated | 21<br>16<br>30<br>33 | 347 |
| **Tree locations** | Location of farm trees | Fields and perhaps home garden<br>Home garden only<br>None cultivated | 40<br>27<br>33 | 349 |
| **Product sales** | Whether tree products sold | Sold products<br>None sold | 15<br>85 | 338 |

Source: Shinyanga District household survey, 2000

### Table 6　Variables found to be significant and relative frequencies

| Resource variables | Brief description | Categories | Relative frequencies | Cases |
|---|---|---|---|---|
| **Farm size** | How large is the household's farm? | 0 – 4 acres<br>5 – 10<br>11 – 20<br>21+ | 18%<br>38<br>29<br>15 | 335 |
| **Land tenure** | Does household own their entire farm? | Private (customary or formal)<br>Partly borrowed or rented<br>Wholly borrowed or rented | 82<br>12<br>6 | 348 |
| **Soil type** | Does farm have *mbuga* clay soil, which is rich, fertile & resists erosion? | At least some *mbuga* soil | 74 | 340 |
| **Household labour** | Total adult worker equivalents in household | 3 or less<br>4 – 6<br>7 or more | 28<br>47<br>25 | 348 |
| **Land – labour ratio** | Number of acres per labour equivalent | < 1<br>1 -2<br>2.5 – 4<br>4 – 10<br>10+ | 12<br>36<br>33<br>16<br>3 | 334 |
| **Labour hired in** | Did household hire in farm labour in past year? | Yes | 21 | 338 |
| **Labour hired out** | Did household hire out farm labour in past year? | Yes | 17 | 349 |
| **Non-farm income** | Does household have a major source of non-farm income, e.g., running a kiosk, working as a traditional healer or teacher? | One or more such source | 17 | 349 |
| **Livestock holdings** | Does household keep livestock? | Own at least some cattle<br>Goats/Sheep/chicken only<br>No livestock | 73<br>22<br>5 | 346 |
| **Market access** | Does household have ready access to a local market? | Remote from markets (>3km)<br>Near to markets (<3 km) | 21<br>79 | 349 |
| **Changes over time in off-farm trees** | Is incidence of off-farm trees seen to be changing by households? | Increasing<br>Roughly stable<br>Decreasing | 0<br>65<br>35 | 349 |

## Table 6   Variables found to be significant and relative frequencies, continued

| Knowledge variables | GK or LK? | Brief description | Categories | Relative frequencies | Cases |
|---|---|---|---|---|---|
| Religion | GK | Religion | Christian or Muslim<br>Pagan / indigenous | 46%<br>54 | 347 |
| Swahili | GK | Swahili language of principal informant | Minimal<br>Intermediate speaking<br>Speak well | 17<br>10<br>73 | 333 |
| Radio | GK | Listens to radio | Don't listen<br>Occasionally<br>Daily / weekly | 44<br>28<br>28 | 349 |
| Extension contact | GK | Contact with HASHI agroforestry outreach | Yes | 22 | 349 |
| Key product: Fuelwood | LK | Cited fuelwood as key tree product of a selected tree | Yes | 67 | 308 |
| Key product: Fruit | LK | Cited fruit as key product of a selected tree | Yes | 16 | 308 |
| Key product: Construction wood | LK | Cited construction wood as key product of a selected tree | Yes | 26 | 308 |
| Key product: Herbal medicine | LK | Cited herbal medicine as key product of a selected tree | Yes | 10 | 308 |
| Key product: Fodder | LK | Cited livestock fodder as key product of a selected tree | Yes | 8 | 308 |
| Soil fertilisation | LK | Cited fertilisation capacity of trees | Yes | 28 | 346 |
| Harm to crops via soil | LK | Suggested trees harm crops via soil effects | Yes | 48 | 346 |
| Harm to crops via shading | LK | Suggested trees harm crops via shading | Yes | 37 | 346 |
| 'Trees make rain' | LK | Cited the traditional belief 'trees make rain' | Yes | 32 | 273 |
| Windbreak / erosion/ microclimate effects | LK | Cited either windbreak, soil erosion, or microclimate effects associated with trees | Yes | 52 | 273 |

| Other variables | Brief description | Categories | Relative frequencies | Cases |
|---|---|---|---|---|
| Kitchen waste | Uses kitchen waste as soil amendment | Yes | 15% | 347 |
| Manure | Uses manure as soil amendment | Yes | 54 | 348 |
| Fallowing | Practices fallowing | Yes | 34 | 347 |
| Crop rotation | Practices crop rotation | Yes | 52 | 336 |
| Intercropping | Practices intercropping | Yes | 86 | 346 |
| Micro-irrigation | Practices micro-irrigation | Yes | 38 | 284 |
| Hybrid seeds | Has purchased hybrid crop seeds, past year | Yes | 32 | 349 |
| Legumes | Grows leguminous crops | Two+<br>One<br>None | 43<br>41<br>16 | 348 |
| Vegetables | Cultivates vegetables | Yes | 45 | 347 |

Source: Shinyanga District household survey, 2000

## Table 7   Explanatory power (i.e., $R^2$) of different versions of the logit models

⇒ Column 1 includes only resource variables such as land, labour and capital
⇒ Column 2 also includes GK variables such as schooling and extension contact
⇒ Column 3 also includes LK variables vis-à-vis tree characteristics

| Regression Model | RVs Only | Add GKVs | Add LKVs | Multiple gain |
|---|---|---|---|---|
| L1, Total number of farm trees | 11.3% | 14.5% | 26.2% | 1.8 |
| L2, Sales of tree products | 12.1 | 17.5 | 21.5 | 1.2 |
| L3, Neither retained nor cultivated farm trees | 11.5 | 15.5 | No change | 0.0 |
| L4, Retained farm trees but did not cultivate them | 8.6 | 15.5 | 26.7 | 1.7 |
| L5, Cultivated exotic fruit trees | 8.1 | No change | 30.4 | 3.8 |
| L6, Planted & watered indig. fruit or non-fruit trees | 1.8 | No change | 12.9 | 7.2 |
| L7, Tree management intensity | 8.6 | 12.0 | 22.6 | 1.9 |
| L8, Type of tree cultivated | 14.8 | 21.0 | 46.6 | 2.2 |
| L9, Location of tree cultivation | 5.8 | 12.8 | 20.8 | 1.6 |

Source: Shinyanga District household survey, 2000

**Table 8   Factors found to be significant in tree management models**

| Factors | Logit 1 | Logit 2 | Logit 3 | Logit 4 | Logit 5 | Logit 6 | Logit 7 | Logit 8 | Logit 9 | Total |
|---|---|---|---|---|---|---|---|---|---|---|
| **Farm size** | +** | − | | +** | −** | | −** | Y** | −* | 7 |
| **Land tenure** | +** | | | | +* | | | | | 2 |
| **Soil type** | | | | | | + | | | | 1 |
| **Labour hired in** | | +** | −* | − | | | +** | Y* | +* | 6 |
| **Labour hired out** | −* | +* | − | | | | +* | | | 4 |
| **Non-farm income** | +* | | | | | | + | | | 2 |
| **Livestock holdings** | | + | −** | | +** | | +* | Y | + | 6 |
| **Changes over time in off-farm trees** | | −** | | − | | | | Y | +* | 4 |
| **Religion** | +** | | −** | −* | + | | +** | Y** | +* | 7 |
| **Swahili** | − | | | | | | | | | 1 |
| **Radio** | | +* | | +** | | | +** | Y* | + | 5 |
| **Extension contact** | + | +** | | −** | | | +** | Y** | +** | 6 |
| **Key product: Fodder** | | | | +** | −* | | −** | Y** | −** | 5 |
| **Key product: Herbal medicine** | | | | −* | | +** | +* | Y** | + | 5 |
| **Key product: Construction wood** | | | | −* | | +** | +** | Y | +** | 5 |
| **Key product: Other** | +* | | | | | | | | | 1 |
| **Soil fertilisation** | | | | −* | | | | Y* | | 2 |
| **Harm to crops via soil** | +* | − | | −** | +* | | | Y** | | 5 |
| **Harm to crops via shading** | +** | +** | | +** | | | | | | 3 |
| **'Trees make rain'** | − | − | | | | | − | | | 3 |
| **Windbreak / soil erosion / microclimate effects** | | + | | | | | | | | 1 |

Source: Shinyanga District household survey, 2000

Key: + and − signify positive and negative regression coefficients, while **, * and no asterisk signify 1%, 5% and 10% significance, respectively.

**Figure 3   Diagram of the associations analysis**

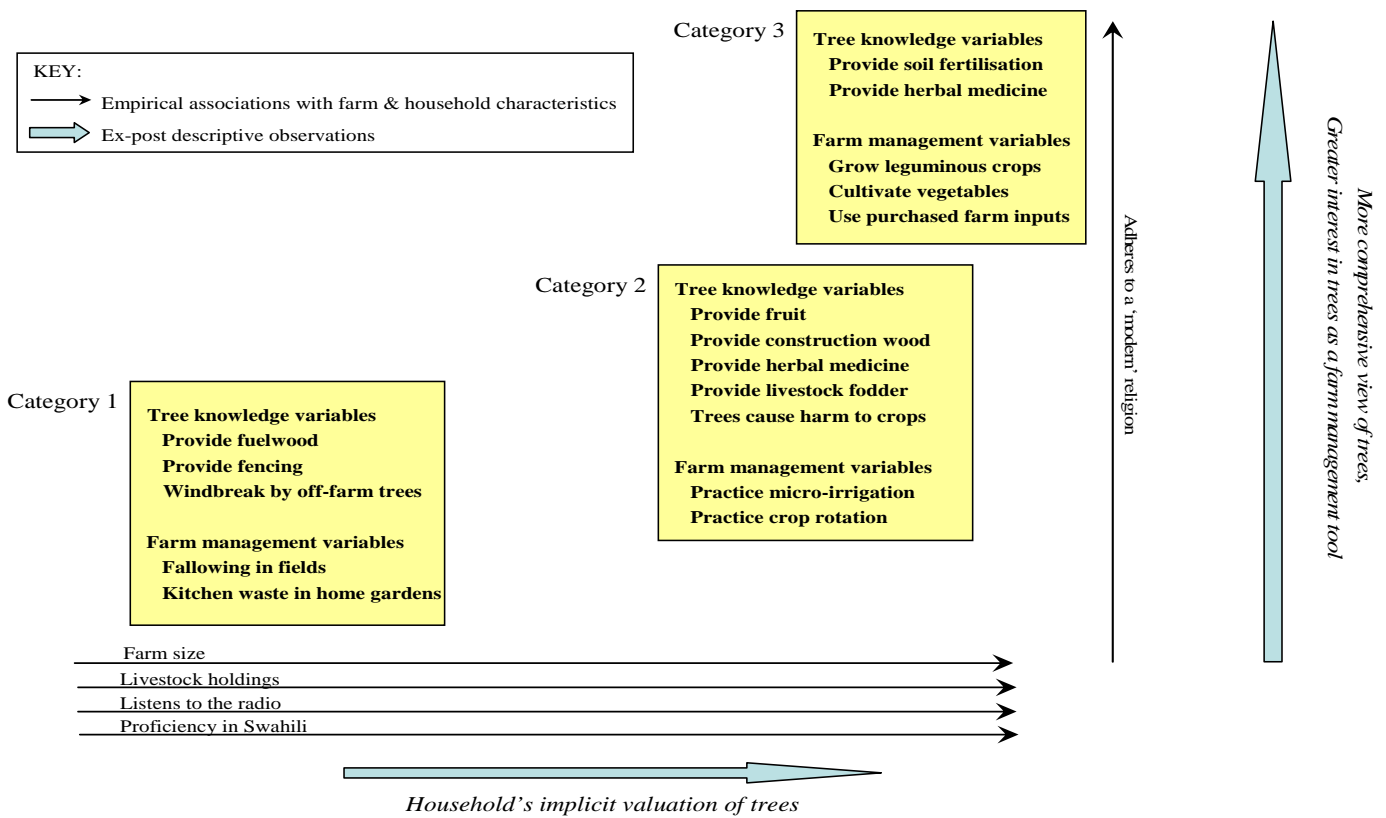

### Table 9A   Given emerging NR scarcity, the demands upon the LK informing NRM are great

| Demands on LK | Elaboration of demands on the adaptive capcity of LK, followed by key questions facing farmers |
|---|---|
| **Context-dependent behaviour of NRs** | Renewable NRs (e.g., soil, pasture, trees, wildlife, fish) reproduce themselves spontaneously via natural regeneration under favourable conditions, yet may do so only slowly and with difficulty where stocks become depleted or conditions are unfavourable.<br>Q:  When will natural regeneration fail to replenish needed local NRs, given current use rates? |
| **Radical changes in optimal NR management** | Where formerly abundant NRs become scarce, maintaining optimal NRM may involve making a radical transition from 'passive' to 'active' management, or from selective exploitation to actively facilitating regenerative processes – e.g., cultivating trees, building up soil organic matter, fish farming.<br>Q:  When must natural regeneration be actively facilitated in order to maintain critical NR stocks? |
| **Economic role of NRs such as trees may be complex** | NRs such as farm trees may provide diverse products and environmental services to rural land managers while also imposing economic costs.  Moreover, trees may interact either symbiotically or competitively with other components of the farm system, e.g., crops, livestock.<br>Q:  When estimating net NR values, which of their diverse costs and benefits should be weighed up? |
| **Missing markets for environmental goods and services** | Many NRs – and associated products and environmental services – are not exchanged in markets, due to their historical abundance and tendency to reproduce spontaneously.  As such, NRs often lack market prices, obliging farmers to estimate their net present value via 'implicit valuation', typically in their heads without clipboards or calculators.  These estimates are critical to determining how NRs are managed, and hence total NR stock levels.<br>Q:  What is the value of an environmental product or service lacking a market price? |
| **Density-dependence of NR values** | In areas where NRs are abundant they are of low value, since any benefits provided are virtually free while any costs imposed may be significant.  For instance, where trees are abundant, benefits such as fuelwood are nearly free while costs such as shading of crops are a major problem.  Yet these same NRs may become valuable when they are scarce, since their costs are low and their benefits are rare and hence precious.<br>Q:  How does the net value of non-marketed NRs change given shifting resource abundance? |

### Table 9B   Reasons for concern that LK may adapt only slowly to emerging NR scarcity

| Constraints on LK | Elaboration of constraints on the adaptation of LK to emerging challenges |
|---|---|
| **NRs as the original 'free lunch'** | NRs tend to reproduce themselves spontaneously via natural regeneration, without need of labour or other inputs.  Natural capital thus differs from other capital, since only in this case is value produced 'for free'.  The trouble is that the 'free' reproduction of resources can encourage users to take these resources for granted, despite the often critical goods and services they provide.  This danger is greatest where NRs were historically abundant. |
| **Settled ideas of 'best practice' and socio-cultural embedding** | Households may hold settled views about how things are done in the local context that are not typically held up to critical scrutiny, whether due to being based on long-established collective historical experience (e.g., "this is how it is done here") or being embedded in local mythology and/or institutions (e.g., "bad spirits live in that type of tree").  Normally, such ideas will be well-suited to the local context.  But where the rural context has changed dramatically, e.g., due to environmental degradation, such distilled experience may prove a poor guide to current NRM challenges. |
| **Provision to communities of inappropriate GK** | Peasant farmers may adopt views ill-suited to their circumstances based on the authority of powerful outsider actors, and their association with success in the popular imagination.  Moreover, exposure to forceful assertions, e.g., regarding the superiority of mono-cropping or removal of trees from fields, may shake farmers' faith in their own judgement. |
| **Adverse reactions against coercive policies** | Rural people may neglect advantageous ideas or ways of thinking due to their association with oppressive outside agents.  One example is British colonial soil conservation policy in the 1950s, which stressed the importance of soil conservation yet led to anti-conservationist views among East African farmers culminating in mass protests (Maack 1996).  Other examples include current policies mandating de-stocking or imposing fines for harvesting products from nature reserves, which seek to safeguard the productivity of pastures and forests yet often inspire resistance. |

### Figure 4   "A clean field is a good field"

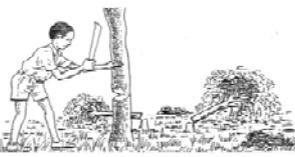
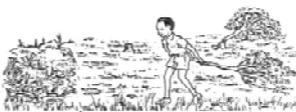
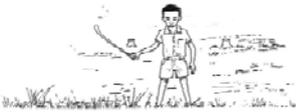
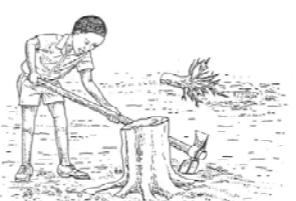
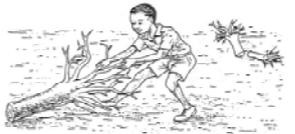
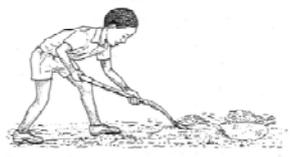
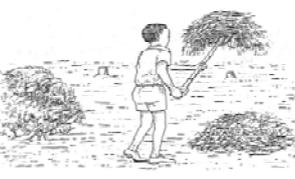
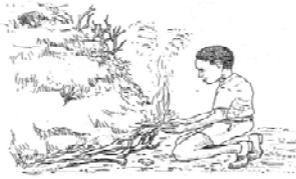

### Table 10   Comparing the three broad NRM strategies available to farmers

|  | A. Exploitation of NRs | B. Cultivation of NRs | C. Stewardship of NRs |
|---|---|---|---|
| **NRM challenge facing farmers** | Accessing and harvesting naturally occurring plants or game | Cultivating selected plant or animal products, relying partly on a favourable local environment | Securing the basis for ongoing, efficient production of targeted products by actively maintaining critical environmental services |
| **General examples** | Hunting, gathering, grazing or logging in 'the bush' | Crop or livestock production, horticulture | Green manure, composting, agroforestry |
| **Tree management example** | Harvesting fuelwood or indigenous fruit from naturally occurring off-farm trees | Cultivating farm trees to obtain a coveted product, such as fruit or construction wood | Cultivating farm trees to bolster the productivity of the farm system, e.g., via erosion control, soil fertilisation, microclimate effects |
| **Situation where strategy most appropriate** | Targeted NRs are ample, such that harvesting does not threaten stocks | Local NRs are either ample or unnecessary, due to ready access to substitute inputs such as chemical fertiliser or irrigation | Local NRs are needed yet scarce, e.g., due to providing critical environmental services or farm inputs for which substitutes are unavailable, e.g., fertiliser, soil moisture |
| **Payoff on inputs to management** | Inputs go to obtaining desired product; payoff is direct (same day) | Inputs go to producing desired product; payoff is fairly direct (same season) | Inputs go to securing basis of production; payoff is indirect (over future seasons) |

### Table 11   The two facets of the knowledge problem: Farm practice, development theory

|  | Theory-practice dichotomy | Farmers' thinking and theory dovetail to mask gaps in LK | Solution to knowledge problem |
|---|---|---|---|
| **Practice of how LK informs NRM on farms** | Evidence presented suggests that farmers' ideas about 'best practice' NRM may be problematic, given emerging NR scarcity | Farmers don't perceive problems with their ideas about 'best practice', since if they did they would amend them | Addressing knowledge gaps so that farmers recognise emerging opportunities |
| **Theory about the LK informing NRM** | Yet theory – and hence development professionals – tend to see LK as unproblematic | Existing theory assumes away problems with LK vis-à-vis NRM, masking any knowledge problems | Amending theory in the specific case of emerging NR scarcity |

### Table 12   Alternative means for peasant farmers to meet key household needs

| Household needs | A. Traditional extensive agriculture | B. Green Revolution agriculture | C. Intensive sustainable agriculture |
|---|---|---|---|
| **Cooking fuel** | Wood from the bush, charcoal, manure | Gas stove, rural electrification, solar oven | Wood from cultivated trees, charcoal |
| **Livestock fodder** | Graze in commons or bush | Improved pasture grasses, feed supplements, silage | Enclosed grasses, cultivated fodder trees |
| **Medicine** | Gather herbal medicines | Pharmacy, clinic, hospital | Cultivated plants and trees provide key herbal medicines; others gathered in the bush |
| **Soil fertilisation** | Fallowing, burn weeds & crop residues | Chemical fertilisers | Manure, green manure, composting, mulching, N-fixing trees, conservation tillage |
| **Watering crops** | Rainfall, choosing soils with good water-holding capacity | Irrigation works to control and increase water supply | Micro-irrigation, raise soil water-holding capacity via increased soil organic matter, grow trees to foster clement microclimate |
| **Pesticide** | Remove vegetation to reduce pest habitat | Chemical pesticides | Integrated pest management, natural pesticides, intercropping |